\documentclass[reprint,aps,prd,nofootinbib,preprintnumbers,twocolumns,showpacs,10pt, superscriptaddress]{revtex4-1}
\usepackage{epsfig}
\usepackage{epstopdf}
\usepackage{hyperref}
\usepackage{amssymb}
\usepackage{amsmath}
\usepackage{amsfonts}
\usepackage{graphicx}
\usepackage{bm}
\usepackage[usenames,dvipsnames]{color}
\usepackage{hyperref}

\def	\be	{\begin{equation}}
\def	\ee	{\end{equation}}
\def	\bqt	{\begin{quote}}
\def	\eqt	{\end{quote}}

\begin{document}

\title{Generalization of the Majorana equation for real spinors}

\author{Gin\'{e}s R.P\'{e}rez Teruel}

\affiliation{Departamento de F\'{i}sica Te\'{o}rica, Universidad de Valencia, Burjassot-46100, Valencia, Spain} 
\email{gines.landau@gmail.com}
\begin{abstract}
\begin{center}
{\bf Abstract}
\end{center}
\noindent
We show that the Dirac equation for real spinors can be naturally decomposed into a system of two first-order relativistic wave equations. The decomposition separates in a transparent way the real and imaginary parts of the Dirac equation by means of two algebraic differential operators, allowing to describe real spinors in any representation of the Dirac matrices maintaining the reality condition
$\tilde{\Psi}=\tilde{\Psi}^{*}$ unaltered. In addition, it is shown that the Majorana wave equation is a particular case of
the relativistic system of equations deduced in this paper. We also briefly discuss how the formalism can be extended to deal with complex (charged) spinors.
\end{abstract}

\maketitle
\section{Introduction}
\label{Introduction} 
\thispagestyle{empty}

\noindent
In spite of the fact that the Dirac and Majorana equations have essentially the same mathematical structure, their physical meaning and interpretation are quite different.The Majorana equation describes neutral fermions that are their own antiparticles, while the Dirac equation is suitable for charged fermions such as the electron or the muon. Although there is no experimental evidence so far of elementary Majorana particles, the presence of Majorana modes has been recently reported in certain superconductor substances \cite{Wilc,Mourik,Nadj,Ho}. On the other hand, there exists an old debate regarding the ultimate nature of neutrinos. Indeed, theoretical physicists widely support the idea that neutrinos are in fact Majorana particles \cite{Bilenky}, namely, real spinorial fields, though this is still being discussed \cite{Czakon,Tho}. If neutrinos are Majorana fermions, the first signal should come from neutrinoless double $\beta$ decay, a process that has not yet been observed \cite{Ago}. 
In any case, the study of the theoretical aspects of real spinors and the possible generalizations of the Majorana equation represent two very interesting research avenues. This is the subject of this work.
Let us begin by writing down the Dirac equation in the usual covariant notation 
\begin{equation}\label{Dirac}
(i\gamma^{\mu}\partial_{\mu}-m)\Psi=0
\end{equation}

Where the Dirac matrices $\gamma^{\mu}$, satisfy the anticommutation rules
\begin{equation}
\gamma^{\mu}\gamma^{\nu}+\gamma^{\nu}\gamma^{\mu}=\{\gamma^{\mu},\gamma^{\nu}\}=2g^{\mu\nu}
\end{equation}
In 1937, Majorana \cite{Maj} inquired whether it might be possible for a spin 1/2 particle to be its own antiparticle, by attempting to find the equation that such an object would satisfy. To get an equation of Dirac's type such as (\ref{Dirac}) (that is, suitable for spin 1/2) but purely real and thus capable of governing a real field, requires matrices that satisfy the Clifford algebra and are purely imaginary \cite{Wilczek}. Majorana found such purely imaginary representation of the Dirac matrices, the so-called Majorana representation. It is important to note that a real fermion can be consistently described by any representation of the gamma matrices, but the Majorana representation turns out to be the most natural. 
In this representation, the reality condition is given by
\begin{equation}\label{real}
\tilde{\Psi}=\tilde{\Psi}^{*}
\end{equation}
Regarding the Dirac matrices, written as tensor products of the usual Pauli matrices $\sigma_{i}$, take the form
\begin{align}
\tilde{\gamma_{0}}=\sigma_{2}\otimes\sigma_{1}\nonumber\\
\tilde{\gamma_{1}}=i\sigma_{1}\otimes 1\nonumber\\
\tilde{\gamma_{2}}=i\sigma_{3}\otimes 1\nonumber\\
\tilde{\gamma_{3}}=i\sigma_{2}\otimes\sigma_{2}
\end{align}
Then, in the Majorana representation the matrices satisfy the property:
\begin{equation}\label{Majorana}
\tilde{\gamma}_{\mu}=-\tilde{\gamma_{\mu}}^{*}
\end{equation}
This turns the Dirac equation into a purely real equation. For any other representation of the Dirac matrices, the reality condition (\ref{real}) is no longer valid and needs to be modified \cite{Palash}. Indeed, since any two representations of the gamma matrices can be related by a similarity transformation involving a unitary matrix, we can write
\begin{equation}
\Psi=U\tilde{\Psi}
\end{equation}

which implies, $\tilde\Psi=U^{\dagger}\Psi$ ; substituting this result in (\ref{real}) we find
\begin{equation}
U^{\dagger}\Psi=(U^{\dagger}\Psi)^{*}
\end{equation}
Or equivalently
\begin{equation}
\Psi=UU^{T}\Psi^{*}
\end{equation}
These are well known features of the standard theory. Nevertheless, in this work we adopt an alternative approach.
In particular, we show that the reality condition (3) in the Majorana representation of a real spinor does
not obligatorily require the use of purely imaginary matrices to get a purely real equation. Indeed, if we want to
obtain a real equation from (1) using complex matrices, this will require in general an explicit decomposition of
the Dirac equation into two other wave equations which are their real and imaginary parts. The paper is organized
as follows. In sec.II we begin with a general analysis of the properties of the Dirac operator, and show
that it can be canonically decomposed into two different parts, one of which is closely related with the Hamiltonian.
Then, we derive two independent first-order wave equations that should satisfy any real spinor. We also
show that these two first-order differential equations contain the Majorana picture as a particular case. Finally,
it is shown how the formalism can be adapted to the case of complex spinors as well.

\section{Decomposition of the Dirac equation}\label{DecompositionDirac}
It is well known that the Dirac equation is in some sense a decomposition (square root) of the Klein-Gordon equation. One naturally wonders whether the Dirac equation, in turn, may be decomposed into other(s) more fundamental equation(s).The question is interesting and, as far as I know, it has already been explored in the literature. To be precise, we should distinguish here between \emph{decomposition} and \emph{factorization}. In this work, we are not interested in the factorization of the Dirac equation into other equations of fractional order employing fractional calculus. This task has been addressed elsewhere, and the interested reader is referred to \cite{Raspin,Raspini,Muslih,Kirchanov} for further information.Rather, we decompose the Dirac equation
into two other differential equations of the same (first) order. The motivation to search for such a decomposition
lies in the study of the real spinors, but we will show later that it can be extended in a natural manner to the
case of complex (charged) spinors as well. This will allow us to deepen in the algebraic structure of the Dirac Lagrangian to discover a particular set of transformations of the global gauge symmetry with very interesting properties.
\subsection{Properties of the general Dirac operator}
Let us start by isolating the general Dirac operator from equation (\ref{Dirac})

\begin{equation}\label{operator}
\hat{O}=i\hat{\gamma}^{\mu}\partial_{\mu}-m\hat{I}=i\hat{\gamma}^{\mu}\partial_{\mu}-m\hat{I}
\end{equation}
Where $\hat{I}$ is the identity operator. Note that such definition
of the Dirac operator does not satisfy the hermiticity
condition:
\begin{align}\label{hermitian}
\hat{O}^{\dagger}&=-i\hat{\gamma}^{\mu\dagger}\partial_{\mu}-m\hat{I}=-i\gamma^{0}\hat{\gamma^{\mu}}\gamma^{0}\partial_{\mu}-m\hat{I}\nonumber\\
&=-\gamma^{0}\Big(i\hat{\gamma^{\mu}}\partial_{\mu}+m\hat{I}\Big)\gamma^{0}
\end{align}
Where we have used the well known property, $\gamma^{\mu\dagger}=\gamma^{0}\gamma^{\mu}\gamma^{0}$. Of course the non-hermiticity of the general Dirac equation is not a problem because one can always construct from it a Hermitian Hamiltonian $\hat{\mathcal{H}}$, which is the meaningful physical operator. In any case, can we express the general Dirac operator as some sort of combination of Hermitian and Antihermitian operators? The answer is yes. Indeed, for any operator $\hat{O} $, it is always possible to write their decomposition in the form \cite{Bohm}
\begin{equation}\label{decomposition}
\hat{O}=\Big(\frac{\hat{O}+\hat{O^{\dagger}}}{2}\Big)+i\Big(\frac{\hat{O}-\hat{O}^{\dagger}}{2i}\Big)=\hat{A}+i\hat{B}
\end{equation}
From this equation, it automatically follows that, $\hat{A}=\hat{A}^{\dagger}$, $\hat{B}=\hat{B}^{\dagger}$.
On the other hand, we have the important property
\begin{equation}\label{square}
\hat{O}\hat{O}^{\dagger}=\Big(\hat{A}+i\hat{B}\Big)\Big(\hat{A}-i\hat{B}\Big)=\hat{A}^{2}+\hat{B}^{2}+i\Big[\hat{B},\hat{A}\Big]=(\Box+m^{2})\hat{I}
\end{equation}
Where $\Box$ is the D'Alembert operator. This means that the ``square" of the general Dirac operator has to agree with the diagonal operator of the Klein-Gordon (KG) equation. Let us apply now this decomposition to the Dirac operator (\ref{operator}). Making use of (\ref{hermitian}), (\ref{decomposition}) and (\ref{square}), we obtain
\small
\begin{align}
\hat{A}&=\frac{1}{2}\Big(i(\gamma^{\mu}-\gamma^{\mu\dagger})\partial_{\mu}-2m\hat{I}\Big)=\frac{1}{2}\Big(i(\gamma^{0}\gamma^{0}\gamma^{\mu}-\gamma^{0}\gamma^{\mu}\gamma^{0})\partial_{\mu}-2m\hat{I}\Big)\nonumber\\
&=\frac{1}{2}\Big(i\gamma^{0}(\gamma^{0}\gamma^{\mu}-\gamma^{\mu}\gamma^{0})\partial_{\mu}-2m\hat{I}\Big)=\gamma^{0}\sigma^{0\mu}\partial_{\mu}-2m\hat{I}
\end{align}
\normalsize
Where we have used that, $(\gamma^{0})^{2}=\hat{I}$, and $\sigma^{\mu\nu}\equiv\frac{i}{2}[\gamma^{\mu},\gamma^{\nu}]$. In the same fashion, for the operator $\hat{B}$ we find
\begin{align}
\hat{B}&=\frac{1}{2}(\gamma^{\mu}+\gamma^{\mu\dagger})\partial_{\mu}=\frac{1}{2}(\gamma^{0}\gamma^{0}\gamma^{\mu}+\gamma^{0}\gamma^{\mu}\gamma^{0})\partial_{\mu}\nonumber\\
&=\frac{\gamma^{0}}{2}\{\gamma^{0},\gamma^{\mu}\}\partial_{\mu}=\gamma^{0}g^{0\mu}\partial_{\mu}=\gamma^{0}\frac{\partial}{\partial t}
\end{align}
Therefore, the imaginary part operator $\hat{B}$ is closely related with the Hamiltonian $\hat{\mathcal{H}}$ of the Dirac equation. The exact algebraic relation among both operators is given by
\begin{equation}
\hat{B}=-i\gamma^{0}\hat{\mathcal{H}}
\end{equation}
Then, the decomposition (\ref{decomposition}) for the Dirac operator turns out to be
\begin{equation}
\hat{O}\equiv\hat{A}+i\hat{B}=\hat{A}+\gamma^{0}\hat{\mathcal{H}}
\end{equation}
Let us conclude the paragraph stressing that the square property (\ref{square}) for the case of the general Dirac operator implies
\begin{equation}\label{squareS}
\hat{O}\hat{O}^{\dagger}=\Big(\hat{A}+\gamma^{0}\hat{\mathcal{H}}\Big)\Big(\hat{A}+\gamma^{0}\hat{\mathcal{H}}\Big)=\hat{A}^{2}+\hat{\mathcal{H}}^{2}+\{\hat{A},\gamma^{0}\hat{\mathcal{H}}\}
\end{equation}
\subsection{Real spinors. Generalization of the Majorana equation}

It is worth noting that there exists at least another possible decomposition of the Dirac operator different from (\ref{decomposition}). Indeed, if one is interested in the explicit separation of the real and imaginary parts of the operator $\hat{O}$, then one should replace in (\ref{decomposition}) the Hermitian conjugation by the complex conjugation, namely

\begin{equation}\label{Realdec}
\hat{O}=\Big(\frac{\hat{O}+\hat{O^{*}}}{2}\Big)+i\Big(\frac{\hat{O}-\hat{O}^{*}}{2i}\Big)=\hat{\mathcal{A}}+i\hat{\mathcal{B}}
\end{equation}
where, $\hat{\mathcal{A}}$, $\hat{\mathcal{B}}$, are now real operators by construction, i.e, $\hat{\mathcal{A}}=\hat{\mathcal{A}^{*}}$, $\hat{\mathcal{B}}=\hat{\mathcal{B}^{*}}$, and notice that are different from those of the Hermitian decomposition, namely, $\hat{\mathcal{A}}\neq\hat{A}$, $\hat{\mathcal{B}}\neq\hat{B}$. 

The decomposition (\ref{Realdec}) is the relevant for our purposes. Indeed, for a real spinor, $\tilde{\Psi}=\tilde{\Psi}^{*}$ the Dirac equation decomposed in such a way will be
\begin{equation}\label{Rdecomposition}
\Big(\hat{\mathcal{A}}+i\hat{\mathcal{B}}\Big)\tilde{\Psi}=\hat{\mathcal{A}}\tilde{\Psi}+i\hat{\mathcal{B}}\tilde{\Psi}=0
\end{equation}
It is important to bear in mind that we cannot identify the real part of the Dirac operator (\ref{operator}) with the mass term, because the gammas are in general complex. Then, in general, $\hat{\mathcal{A}}\neq -m\hat{I}$, and such trivial identification only takes place for a purely real representation of the Dirac matrices. Indeed, apart of the real part of the mass parameter, $\hat{\mathcal{A}}$ will also include in general another contribution where the imaginary part of the gammas plays a role. Since, $\hat{\mathcal{A}}$, $\hat{\mathcal{B}}$, $\tilde{\Psi}$ are all real valued, equation (\ref{Rdecomposition}) is satisfied if and only if
\begin{align}
\hat{\mathcal{A}}\tilde{\Psi}=0\\
\hat{\mathcal{B}}\tilde{\Psi}=0
\end{align}
or equivalently, in the standard notation
\begin{equation}\label{wavereall}
\Big(i\tau^{\mu}\partial_{\mu}-m\Big)\tilde{\Psi}=0
\end{equation}
\begin{equation}\label{eta}
\eta^{\mu}\partial_{\mu}\tilde{\Psi}=0
\end{equation}
Where we have defined two auxiliary matrices $\hat{\tau}$, $\hat{\eta}$ as
\begin{equation}\label{tau}
\hat{\tau}^{\mu}\equiv\frac{1}{2}\Big(\hat{\gamma}^{\mu}-\hat{\gamma}^{*\mu}\Big)
\end{equation}
\begin{equation}\label{etaa}
\hat{\eta}^{\mu}\equiv\frac{1}{2}\Big(\hat{\gamma}^{\mu}+\hat{\gamma}^{*\mu}\Big)
\end{equation}
From the above expressions it automatically follows that
\begin{equation}
\hat{\tau^{\mu}}+\hat{\eta}^{\mu}=\hat{\gamma}^{\mu}
\end{equation}
Eqs. (\ref{wavereall}-\ref{eta}) represent two first-order real eigenvalue equations that should be satisfied simultaneously by any real spinor. Since $\hat{\tau}$ are purely imaginary matrices by construction, while $\hat{\eta}$ are real, the reality of both wave equations is guaranteed. It is easy to demonstrate that the standard Majorana equation is a particular case of this system. Indeed, notice that if we choose purely imaginary Dirac matrices (\ref{Majorana}), i.e $\hat{\gamma}^{\mu}=-\hat{\gamma}^{*\mu}$,the auxiliary matrices (\ref{tau}-\ref{etaa}) reduce to, $\hat{\tau}^{\mu}=\hat{\gamma}^{\mu}$, and, $\hat{\eta}^{\mu}=0$. As a consequence, equation (\ref{wavereall}) reduces to the real Majorana wave equation, while the other wave equation (\ref{eta}), vanishes identically.  One advantage of this approach lies in the fact that it allows to work in any representation of the gamma matrices maintaining the reality condition (\ref{real}) unaltered. The purely real representation of the Dirac matrices is also included in our picture, but the combination of a real spinor with a real representation of the gammas is consistent only with massless particles. Indeed, when $\hat{\gamma}^{\mu}=\hat{\gamma}^{*\mu}$, then $\hat{\tau}^{\mu}=0$, as a consequence Eq. (\ref{wavereall}) sets to zero the mass eigenvalue, i.e, $m=0$, and therefore only the second wave equation (\ref{eta}), which corresponds to a massless particle, will survive.
On the other hand, the real field $\tilde{\Psi}$ can be written in terms of its Fourier expansion as

\begin{equation}\label{Fourier}
\tilde{\Psi}=\sum_{s}\int_{p}\Big(a_{s}\tilde{u}_{s}(p)e^{-ip\cdot x}+{a}_{s}^{\dagger}(p)\tilde{u}_{s}^{*}(p)e^{ip\cdot x}\Big)
\end{equation}
Notice that the summation only involves two values of $s$ due to the fact that the four basis spinors consist on two Spinors $\tilde{u}_{s}(p)$ and their complex conjugates. Substituting the expansion (\ref{Fourier}) into the wave equations (\ref{wavereal1}-\ref{eta}), we get
\begin{equation}\label{dec}
\Big(\tau^{\mu}p_{\mu}-m\Big)\tilde{u}_{s}(p)=0
\end{equation}
\begin{equation}\label{31}
\Big(\tau^{\mu}p_{\mu}+m\Big)\tilde{u}_{s}^{*}(p)=0
\end{equation}
\\
since $\tau^{\mu}=-\tau^{*\mu}$, it is easy to see that both equations are the complex conjugate of each other. Furthermore, we have
\begin{equation}\label{dec2}
\eta^{\mu}p_{\mu}\tilde{u}_{s}(p)=0
\end{equation}
\begin{equation}\label{33}
\eta^{\mu}p_{\mu}\tilde{u}_{s}^{*}(p)=0
\end{equation}
\\
As a consistency proof, note that if we sum the two equations satisfied by $\tilde{u}_{s}(p)$, i.e. Eqs. (\ref{dec}) and (\ref{dec2}), the result obtained is nothing but the general Dirac equation (with complex mass) in momentum space
\begin{align}
\Big(\gamma^{\mu}p_{\mu}-m\Big)\tilde{u}_{s}(p)=0
\end{align}
The same can be stated for the corresponding complex conjugates basis spinors $\tilde{u}_{s}^{*}(p)$. Indeed, the addition of Eqs. (\ref{31}-\ref{33}) provides
\begin{equation}
\Big(\gamma^{\mu}p_{\mu}+m\Big)\tilde{u}_{s}^{*}(p)=0
\end{equation}
A natural question arises: which is the Lagrangian density that allows one to derive the two independent eigenvalue equations of our picture? It is straightforward to verify that such a Lagrangian can be expressed as
\begin{equation}
\mathcal{L}=\overline{\tilde{\Psi}}\Big(\hat{\mathcal{A}}+i\hat{\mathcal{B}}\Big)\tilde{\Psi}+h.c.
\end{equation}
where the standard abbreviation $h.c.$ stands for the Hermitian conjugate in order to guarantee Hermiticity of the Lagrangian density.
\subsection{Lorentz covariance}
Now we want to raise the issue of the Lorentz covariance of the wave equations (\ref{wavereall})-\ref{eta}). It is not generally well appreciated that, when we get to the Dirac equation or similar constructions, it is either possible to view the gamma matrices $\gamma_{\mu}$ as constants, invariant under Lorentz transformations and view the field $\Psi$ to transform as a spinor under Lorentz transformations, or you can view $\gamma_{\mu}$ as a matrix-valued 4-vector, which transforms as a vector under Lorentz transformations, and view $\Psi$ as a set of 4 Lorentz scalars. The two approaches are mathematically equivalent. Almost all treatments of the Dirac equation view $\Psi$ as a Lorentz spinor and $\gamma^{\mu}$ as 4 constant matrices. However, the other way of doing it is also a legitimate approach, see for instance (\cite{Nikolic}). The combination $\overline{\Psi}\gamma^{\mu}\Psi$ gives the same result in either way of doing it.  Indeed, both pictures are equivalent in the same fashion that in non-relativistic quantum mechanics are equivalent the ``Schrodinger picture" and the ``Heisenberg picture". Here, to show the Lorentz invariance of the wave equations we will adopt the latter approach, namely, we will deal with $\gamma_{\mu}$ as a matrix-valued 4-vector, which transforms as a vector under Lorentz transformations, and view $\Psi$ as a set of 4 Lorentz scalars. Once accepted this, the Lorentz covariance of (\ref{wavereall})-\ref{eta}) is quite trivial. However, let us present the proof for completeness purposes. Under a Lorentz transformation, we have therefore $\Psi^{\prime}=\Psi$, and
\begin{equation}
\gamma^{\mu}=\Lambda^{\mu}_{\nu}\gamma^{\prime\nu}
\end{equation}
Since the Lorentz matrices $\Lambda$ are real, the relations satisfied by the auxiliary matrices defined by (\ref{tau}-\ref{etaa}) are not destroyed under a Lorentz transformation, namely
\begin{align}
\hat{\tau}^{\prime\mu}&=\frac{1}{2}\Big(\hat{\gamma}^{\prime\mu}-\hat{\gamma}^{\prime *\mu}\Big)=\frac{1}{2}\Big((\Lambda^{-1})^{\mu}_{\nu}\hat{\gamma}^{\nu}-(\Lambda^{-1})^{\mu}_{\nu}\hat{\gamma}^{*\nu}\Big)\nonumber\\
&=(\Lambda^{-1})^{\mu}_{\nu}\frac{1}{2}\Big(\hat{\gamma}^{\nu}-\hat{\gamma}^{*\nu}\Big)=(\Lambda^{-1})^{\mu}_{\nu}\hat{\tau}^{\nu}
\end{align}
With these results the proof of the Lorentz covariance of (\ref{wavereall}) is immediate
\small
\begin{align}
\Big(\tau^{\prime\mu}\partial^{\prime}_{\mu}-m\Big)\Psi^{\prime}&=\Big((\Lambda^{-1})^{\mu}_{\nu}\tau^{\nu}\Lambda^{\alpha}_{\mu}\partial_{\alpha}-m\Big)\Psi\nonumber\\
&=\Big((\Lambda^{-1})^{\mu}_{\nu}\Lambda^{\alpha}_{\mu}\tau^{\nu}\partial_{\alpha}-m\Big)\Psi=\Big(\delta^{\alpha}_{\nu}\tau^{\nu}\partial_{\alpha}-m\Big)\Psi\nonumber\\
&=\Big(\tau^{\alpha}\partial_{\alpha}-m\Big)\Psi
\end{align}
\normalsize
We omit the proof of the covariance of Eq. (\ref{eta}), since the methodology is analogue.
\subsection{Complex spinors. Lagrangians densities and exchange symmetry.}
In this subsection we address the natural generalization of the approach to deal also with complex spinorial fields. Let us assume the following decomposition for a complex (charged) four-spinor
\begin{equation}
\Psi=\phi+i\varphi
\end{equation}
where, $\phi$, $\varphi$ are a pair of real spinorial fields. On the other hand, the field $\Psi$ should satisfy the Dirac equation, which we have shown that can be decomposed according to Eq. (\ref{Rdecomposition}), namely
\begin{equation}
\Big(\hat{\mathcal{A}}+i\hat{\mathcal{B}}\Big)\Psi=\Big(\hat{\mathcal{A}}+i\hat{\mathcal{B}}\Big)\Big(\phi+i\varphi\Big)=\hat{\mathcal{A}}\phi-\hat{\mathcal{B}}\varphi+i\Big(\hat{\mathcal{B}}\phi+\hat{\mathcal{A}}\varphi\Big)=0
\end{equation}
Since all the objects that appear in the last equation are real-valued, the equation implies the following system
\begin{align}\label{system}
\hat{\mathcal{A}}\phi=\hat{\mathcal{B}}\varphi\\
\hat{\mathcal{B}}\phi=-\hat{\mathcal{A}}\varphi
\end{align}
This is a system of two first-order coupled differential equations. Making explicit the algebraic expressions of the operators $\tilde{\mathcal{A}}$, $\tilde{\mathcal{B}}$ we have
\begin{align}
\Big(i\tau^{\mu}\partial_{\mu}-m\Big)\phi=\eta^{\mu}\partial_{\mu}\varphi\\
\eta^{\mu}\partial_{\mu}\phi=\Big(-i\tau^{\mu}\partial_{\mu}+m\Big)\varphi
\end{align}
It is worth noting that in situations where $\varphi=0$, or $\phi=0$, the system decouples and reduces to the eigenvalue system of Eqs.(\ref{wavereall}-\ref{eta}) identically. This means that our approach can also be adapted to deal with complex (charged) spinors in a natural fashion. Then, our picture is not aimed to substitute the current formulation of the Dirac theory, rather, it can be useful to complement it, allowing a broader and more general view on the subject. In this sense, it is also interesting to stress that the general Dirac Lagrangian density can be written in the following form
\small
\begin{align}\label{Lagrangian}
\mathcal{L}_{D}&=\overline{\phi}\hat{\mathcal{A}}\phi+\overline{\varphi}\hat{\mathcal{A}}\varphi+\overline{\varphi}\hat{\mathcal{B}}\phi-\overline{\phi}\hat{\mathcal{B}}\varphi+i(\overline{\phi}\hat{\mathcal{B}}\phi+\overline{\varphi}\hat{\mathcal{B}}\varphi+\overline{\phi}\hat{\mathcal{A}}\varphi-\overline{\varphi}\hat{\mathcal{A}}\phi)\nonumber\\
&=\Big(\overline{\phi}-i\overline{\varphi}\Big)\Big(\hat{\mathcal{A}}+i\hat{\mathcal{B}}\Big)\Big(\phi+i\varphi\Big)+h.c.
\end{align}
\normalsize
The system (\ref{system}) is therefore derived by application of the corresponding E-L equations with respect to the variables $\overline{\varphi}$, $\overline{\phi}$. Furthermore, such a decomposition allows us to realize that the Dirac Lagrangian is not algebraically irreducible and turns out to be the addition of two pieces $\mathcal{L}_{D}=\mathcal{L}_{1}+\mathcal{L}_{2}$, each of which can also independently generate the system (\ref{system}) once applied their respective variational equations. These variational equations read $\frac{\partial\mathcal{L}_{1}}{\partial\overline{\varphi}}=0$ ,$\frac{\partial\mathcal{L}_{2}}{\partial\overline{\phi}}=0$, where the pair of Lagrangians $\mathcal{L}_{1}$, $\mathcal{L}_{2}$, are given by
\begin{equation}
\mathcal{L}_{1}\equiv\overline{\varphi}\hat{\mathcal{A}}\varphi+\overline{\varphi}\hat{\mathcal{B}}\phi+i\Big(\overline\varphi\hat{\mathcal{B}}\varphi-\overline{\varphi}\hat{\mathcal{A}}\phi\Big)+h.c.
\end{equation}
\begin{equation}
\mathcal{L}_{2}\equiv\overline{\phi}\hat{\mathcal{A}}\phi-\overline{\phi}\hat{\mathcal{B}}\varphi+i\Big(\overline\phi\hat{\mathcal{B}}\phi+\overline\phi\hat{\mathcal{A}}\varphi\Big)+h.c.
\end{equation}\\
Regarding the invariance properties, note that the full Lagrangian (\ref{Lagrangian}) is invariant under a rotation of angle $\theta$ in the complex plane, namely, a transformation of the type
\begin{equation}\label{gauge}
\begin{pmatrix}
\phi^{\prime}\\
\varphi^{\prime}\\
\end{pmatrix}
=
\begin{pmatrix}
\cos\theta & -\sin\theta\\
\sin\theta & \cos\theta\\
\end{pmatrix}
\begin{pmatrix}
\phi \\
\varphi\\
\end{pmatrix}
\end{equation}
This transformation express the well known global gauge invariance of the Dirac theory. Indeed, note that (\ref{gauge}) is equivalent to the phase transformation $\Psi\rightarrow e^{i\theta}\Psi$. It is worth noting that the two building blocks Lagrangians $\mathcal{L}_{1}$, $\mathcal{L}_{2}$ are also invariant under the same type of transformation, $\phi\rightarrow e^{i\theta}\phi$, $\varphi\rightarrow e^{i\theta}\varphi$. This is evident if we rewrite both Lagrangians in the compact form:
\begin{equation}
\mathcal{L}_{1}=\overline{\varphi}\Big(\hat{\mathcal{A}}+i\hat{\mathcal{B}}\Big)\varphi+\overline{\varphi}\Big(\hat{\mathcal{B}}-i\hat{\mathcal{A}}\Big)\phi+h.c.
\end{equation}
\begin{equation}
\mathcal{L}_{2}=\overline{\phi}\Big(\hat{\mathcal{A}}+i\hat{\mathcal{B}}\Big)\phi+\overline{\phi}\Big(-\hat{\mathcal{B}}+i\hat{\mathcal{A}}\Big)\varphi+h.c.
\end{equation}
Interestingly, there exists an exchange symmetry between both Lagrangians, namely, for certain values of the rotation angle $\theta$ of the gauge transformation (\ref{gauge}), then $\mathcal{L}_{1}$ turns into $\mathcal{L}_{2}$, and vice versa. For instance, let us consider the transformation
\begin{equation}\label{gauge2}
\begin{pmatrix}
\phi^{\prime}\\
\varphi^{\prime}\\
\end{pmatrix}
=
\begin{pmatrix}
0 & 1\\
-1 & 0\\
\end{pmatrix}
\begin{pmatrix}
\phi \\
\varphi\\
\end{pmatrix}
\end{equation}
This transformation is a particular case of (\ref{gauge}) for $\theta=-\pi/2$. Indeed, for such a value of the angle the fields transform as, $\phi\rightarrow\varphi$ , $\varphi\rightarrow-\phi$, which implies, $\mathcal{L}_{1}\rightarrow\mathcal{L}_{2}$, $\mathcal{L}_{2}\rightarrow \mathcal{L}_{1}$. Indeed, note that
\small
\begin{equation}
\mathcal{L}_{1}^{\prime}=\overline{\varphi}^{\prime}\Big(\hat{\mathcal{A}}+i\hat{\mathcal{B}}\Big)\varphi^{\prime}+\overline{\varphi}^{\prime}\Big(\hat{\mathcal{B}}-i\hat{\mathcal{A}}\Big)\phi^{\prime}=\overline{\phi}\Big(\hat{\mathcal{A}}+i\hat{\mathcal{B}}\Big)\phi+\overline{\phi}\Big(-\hat{\mathcal{B}}+i\hat{\mathcal{A}}\Big)\varphi
\end{equation}
\normalsize
Then, the transformation only exchanges the values of the building blocks of the Dirac Lagrangian, but the Dirac Lagrangian itself remains unchanged because corresponds to the addition of both, i.e, $\mathcal{L}_{D}=\mathcal{L}_{1}+\mathcal{L}_{2}$, and only rearranges terms permuting $\mathcal{L}_{1}$ by $\mathcal{L}_{2}$. In general, one can verify that the Lagrangians $\mathcal{L}_{1}$, $\mathcal{L}_{2}$ permute for values of the rotation angle such that $\theta=(2k+1)\pi/2$, $k\in \mathbb{Z}$
\section{FINAL REMARKS}
In this work, we have shown that the Majorana equation for real spinors emerges as a particular case of a system of two first-order differential wave equations. These two wave equations are, in turn, a particular case of another more general system of two first-order coupled wave equations that works for complex (charged) spinors. Indeed, when $\phi=0$, or $\varphi=0$, the system decouples and
collapses into a purely real system describing a real field. In addition, we have demonstrated how this description
is possible due to a canonical decomposition of the Dirac operator, which allows to separate in a transparent way
the real and imaginary parts of the Dirac operator for an arbitrary representation of the Dirac matrices. We have also provided an alternative formulation of the general Dirac Lagrangian formalism, which allowed us to realize that the Dirac Lagrangian $\mathcal{L}_{D}$ is not algebraically irreducible, rather, it turns out to be the combination of two other pieces $\mathcal{L}_{1}$ and $\mathcal{L}_{2}$ each of which is enough to generate by itself the correct wave equations under application of the E-L variational method. Besides, these two pieces are linked by an interesting symmetry, a particular case of the global gauge transformation for $\theta=(2k+1)\pi/2$, $k\in \mathbb{Z}$, under which the Lagrangians permute their values, maintaining the full Dirac Lagrangian unaltered.\\

In the last years, further developments have been paved in the literature, concerning new kinds of spinors and their applications. In particular, from the first quantised point of view the so called Elko spinors \cite{Elko} may play an important role in the dark matter description. A summarised review about the regular and singular spinors, wherein Majorana, Elko, Dirac, Weyl, flagpoles, flag-dipoles, and so on, spinors reside can be found in \cite{DaRocha1}. In Ref. \cite{Cavalcanti} there is table with an algebraic classification of all spinors. The classes 1, 2, and 3 in the table can all describe real spinors. However, it will be interesting to determine exactly to which class belong the spinors that we have discussed in this paper. This important question will be addressed in a future work.


\begin{thebibliography}{99}
\bibitem{Wilc} Wilczek, Frank. Quantum Physics:Majorana Modes Materialize. Nature 486.7402 (2012): 195–197.
\bibitem{Mourik} Mourik, V. et al. Signatures of Majorana fermions in hybrid superconductor-semiconductor nanowire devices. Science 336, 1003–1007 (2012).
\bibitem{Nadj} Nadj-Perge S. et al. Observation of majorana fermions in ferromagnetic atomic chains on a superconductor, Science 346, (2014)
\bibitem{Ho} Ho-Yin Hui et al. Majorana fermions in ferromagnetic chains on the surface of bulk spin-orbit coupled s-wave superconductors, Sci. Rep. 5, 8880 (2015)
\bibitem{Bilenky} See for instance: S.M. Bilenky, C. Giunti, W. Grimus, hep-ph/9812260;
G. Altarelli, F. Feruglio, hep-ph/9905536; P. Fisher, B.Kayser, K.S. McFarland,
hep-ph/9906244; Europhysics Neutrino Oscillation Workshop, hepph/9906251;
R.D. Peccei, hep-ph/9906509; I. Ellis, hep-ph/9907458;
\bibitem{Czakon} M.Czakon et al. Are Neutrinos Dirac or Majorana particles? Acta Phys. Polon. B 30:3121-3138,1999
\bibitem{Tho} Thomas D.Gutierrez; Distinguishing between Dirac and Majorana neutrinos with two-particle interferometry,Phys. Rev. Lett. 96, 121802 (2006)
\bibitem{Ago} M. Agostini et al. Results on Neutrinoless Double-$\beta$ Decay of Ge76 from Phase I of the GERDA Experiment, Phys. Rev. Lett. 111, 122503 
\bibitem{Maj} E. Majorana; Symmetrical Theory of Electrons and Positrons, Nuovo Cim. {\bf{14}} (1937) 171 
\bibitem{Wilczek} F. Wilczek; Majorana returns, Nature Physics 5, 614 - 618 (2009) 
\bibitem{Palash} Palash B. Pal, Dirac, Majorana and Weyl fermions, Am. J. Phys. {\bf{79}}, 485 (2011)
\bibitem{Raspin} A. Raspini, Dirac Equation with Fractional Derivatives of Order 2/3, Fizika B, vol. 9, Issue 1, pp.49-54
\bibitem{Raspini} A. Raspini, 2001 Phys. Scr. {\bf{64}} 20
\bibitem{Muslih} Sami I Muslih et al 2010 J. Phys. A: Math. Theor. {\bf{43}} 055203
\bibitem{Kirchanov} V. S. Kirchanov, The Dirac equation in the fractional calculus, Russ Phys J. 56(9) (2013)
\bibitem{Bohm} D.Bohm, D, Quantum theory, Prentice-Hall, Inc., New York,. 1951,pg 188-189
\bibitem{Nikolic}Nikolic, H, How (not) to teach Lorentz covariance of the Dirac equation, Eur. J. Phys. 35, 035003 (2014)
\bibitem{Elko}D.V. Ahluwalia and D. Grumiller. Spin-Half Fermions with Mass Dimension One: Theory, Phenomenology, and Dark Matter. Journal of Cosmology and Astroparticle Physics, 2005:012, 2005.
\bibitem{DaRocha1} J.M. Hoff Da Silva, R. Da Rocha, Unfolding Physics from the Algebraic Classification of Spinor Fields,  Phys. Lett. B 718 (2013) 1519
\bibitem{Cavalcanti} R.T. Cavalcanti, Classification of Singular Spinor Fields and Other Mass Dimension One
Fermions, Int. J. Mod. Phys. D 23 (2014) 14, 1444002
\end{thebibliography}
\end{document}